
\magnification=\magstep1
\vsize=51true pc
\hsize=36true pc
\baselineskip=14pt
\overfullrule=0pt

\rightline{hep-th/9210137}
\rightline{IASSNS-HEP-92/76}
\rightline{October 1992}

\vskip .2in

\centerline{\bf  Actions for Integrable Systems and}
\centerline{{\bf Deformed Conformal Theories}\footnote*{Based on a talk
   given at the NSERC-CAP Workshop, {\it Quantum Groups, Integrable Models
   and Statistical Systems}, Kingston, Ontario, Canada, July 1992}}

\vskip .3in

\centerline{Jeremy Schiff}
\smallskip
\centerline{\it School of Natural Science, Institute For Advanced Study}
\centerline{\it Olden Lane, Princeton, NJ 08540}

\vskip .3in

\centerline{\bf Abstract}
{\narrower\smallskip
\noindent I report on work on
a Lagrangian formulation for the simplest 1+1 dimensional
integrable hierarchies. This formulation makes the relationship between
conformal field theories and (quantized) 1+1 dimensional integrable
hierarchies very clear.\smallskip}

\vskip .5in

\noindent{\bf 1. Introduction}
\smallskip
It is a not widely appreciated fact that at least some (1+1) dimensional
integrable hierarchies (of KdV type), in their ``second'' hamiltonian
formulation, can be derived from an action principle$^1$.
Interestingly, as I will show later, {\it the same action, considered as
a functional of different sets of fields, can give rise to different
``gauge equivalent'' integrable hierarchies}$^2$.
But the main merit of this
Lagrangian approach to integrable systems is that when we quantize these
theories in the obvious way, we see the relationship between (deformed)
conformal field theories and quantized integrable systems emerge naturally.
We need an action for integrable systems in their ``second'' hamiltonian
formulation, because it is the ``second'' Poisson bracket algebras of
integrable systems that are related to the operator algebras of conformal
field theory$^3$.

I will focus here on the KdV action, summarizing the results of ref.4,
but I will also give an action for NLS$^5$
hierarchy. In section 2, after presenting some results in classical KdV
theory, including  an explanation of the
notion of the ``gauge equivalence class'' of the KdV equation, I
give the KdV action. In section 3 I show how to quantize the theory
defined by the KdV action, to obtain  the usual notions of ``quantum
KdV'' and ``quantum MKdV''; we also obtain very naturally
the result of Zamolodchikov$^6$, that the quantum KdV hamiltonians are
conserved quantities in a certain deformation of the minimal conformal
models. Similar treatment of the NLS action, given in section 4,
reveals the result that the quantum NLS hamiltonians are conserved
quantities in a certain deformation of the parafermion
and $SL(2)/U(1)$ coset models.

\vskip.2in

\noindent{\bf 2. Classical KdV Theory and the KdV Action}
\smallskip

The meaning of the statement ``the KdV equation is gauge equivalent to the
MKdV equation'' is that
via the ``Miura map'' $u=j_x-{1\over 2}j^2$,
a) a solution to the MKdV equation
$$ j_t=j_{xxx}-{\textstyle{3\over 2}}j^2j_x     \eqno{(1)}$$
generates a solution of the KdV equation
$$ u_t=u_{xxx}+3uu_x                            \eqno{(2)}$$
and b) the ``second'' Poisson bracket structure of the MKdV equation$^7$
$$ \{j(x),j(x')\} = \partial_x\delta(x-x')      \eqno{(3)}$$
induces the ``second'' Poisson bracket structure of the KdV equation
$$ \{u(x),u(x')\} = (\partial_x^3 + u(x)\partial_x + \partial_x u(x))
                    \delta(x-x')                \eqno{(4)}$$
Less well known, but of fundamental importance$^8$, is the fact that via
the map $j=q_{xx}/q_x$, a) a solution of the Ur-KdV equation
$$ q_t=q_{xxx}-{\textstyle{3\over 2}}q_{xx}^2q_x^{-1}     \eqno{(5)}$$
generates a solution of MKdV, and b) the Poisson bracket structure
$$ \{q(x),q(x')\} = \partial_x^{-1}q_x\partial_x^{-1}q_x\partial_x^{-1}
                    \delta(x-x')                \eqno{(6)}$$
induces the Poisson bracket structure (3). As recognized
and explained by Wilson$^8$, Eqs.(5) and (6) are invariant under M\"obius
transformations
$$ q \rightarrow {{aq+b}\over{cq+d}} ~~~~~~~~~~~~~ ad-bc=1  \eqno{(7)}$$
It follows that if $f$ is some functional of $q$ invariant under
some subgroup of the M\"obius transformations, then Eq.(5) will imply
some KdV-type equation for $f$ and the bracket (6) will imply some bracket
for $f$. Examples of such $f$'s are $j$, which is invariant under the
$c=0$ subgroup of M\"obius transformations, and $u=q_{xxx}q_x^{-1}-
{3\over 2}q_{xx}^2q_x^{-2}$,
which is invariant under the full group of M\"obius transformations.
Another such $f$ is $\tilde{j}\equiv q_{xx}q_x^{-1}-2q_xq^{-1}$,
which is invariant
under the $b=0$ subgroup; $\tilde{j}$ satisfies MKdV\footnote*{
So there are two distinct maps from Ur-KdV to MKdV.}, satisfies the same
brackets as $j$, and
$u=\tilde{j}_x-{1\over 2}\tilde{j}^2$. We can also take $f$'s that are
invariant under one parameter subgroups of the M\"obius
transformations, such as $h\equiv\ln q_x$ (invariant under $a=d=1$, $c=0$),
$\tilde{h}\equiv\ln(q_x/q^2)$ (invariant under $a=d=1$, $b=0$), and
$\eta\equiv q_x/q$ (invariant under $b=c=0$). All this is explained in ref.8.
Note that we could have written any function of $q_x$ instead of $h$ above;
we have chosen $h$ and $\tilde{h}$ so that $j=h_x$ and
$\tilde{j}=\tilde{h}_x$. The complete set of equations obtained from Ur-KdV
in this way is what I call the gauge equivalence class of KdV; I
should point out that this notion is usually introduced via a
zero-curvature formulation, but I will have no need for this here.

Consider now the following action:
$$ \eqalign{ S  &=S_0+H \cr
             S_0&= -{c\over{48\pi}}\int dxdt~q_{xt}q_{xx}q_{x}^{-2} \cr
             H  &=\int dxdt~p[u] } \eqno{(8)}$$
Here $c$ is a constant and
$p[u]$ is some function of $u$ and its $x$-derivatives.
$S_0$ is the ``geometric Virasoro action'' of Polyakov, Bershadsky and
Ooguri and others$^9$, which is invariant under M\"obius transformations (7),
as is $H$. $H$ has no time derivatives in it, so the Poisson brackets
are determined purely by $S_0$; on the other hand  since $S_0$
is first order in time derivatives it will give no contribution to the
hamiltonian, which is therefore  $H$. The Poisson brackets determined
by $S_0$ are exactly those in Eq.(6) multiplied by $24\pi/c$. We can
write
$$ S_0= -{c\over{48\pi}}\int dxdt~h_xh_t  \eqno{(9)}$$
so $S$ can also be considered as an action for $h$. Treating $S$ as an
action for $q$ the equation of motion is found to be
$$ u_t=-{{24\pi}\over c} (\partial_x^3 + u(x)\partial_x + \partial_x u(x))
            {{\delta p}\over{\delta u}} \eqno{(10)}$$
Here ${{\delta p}/{\delta u}}$ is defined by $\delta H = \int dxdt~
({\delta p}/{\delta u})\delta u$. Treating $S$ as an action for $h$ the
equation of motion is
$$ j_t=-{{24\pi}\over c} \partial_x(\partial_x+j)
            {{\delta p}\over{\delta u}} \eqno{(11)}$$
where here we understand that we should write ${\delta p}/{\delta u}$
in terms of $j$. From (10) and (11) it is clear that if we choose
$$ p[u]=\sum_{n=1}^{\infty} \lambda_np_n[u]  \eqno{(12)}$$
where the $\lambda_n$'s are constants and the $p_n$'s are the densities
of the conserved quantities of the KdV equation,
$$ \eqalign{ p_1[u]&=u\cr
             p_2[u]&={\textstyle{1\over 2}}u^2\cr
             p_3[u]&={\textstyle{1\over 2}}(u^3-u_x^2) \cr
                   &\vdots}\eqno{(13)}$$
then (10) will give an arbitrary equation in the KdV hierarchy and
(11) an arbitrary equation in the MKdV hierarchy. Note that by treating
$S$ as a non-local functional of $u$ we can also obtain an arbitrary
equation in the Ur-KdV hierarchy from $S$.$^4$

In the last paragraph we pulled the KdV conserved quantities out of a hat.
In fact we could have chosen $p[u]$ as in Eq.(12) with the $p_n$'s {\it any}
set of densities such that the quantities $I_n=\int dx~p_n[u]$
mutually commute
under the bracket (4). This would have given a different integrable hierarchy.
I am not aware of a classification of all possible
sets of $p_n$'s. But when we write the $p_n$'s of the KdV hierarchy in
terms of $h$ we find that the $I_n$'s commute with both $I_+=\int dx~e^h$
and $I_-=\int dx~e^{-h}$ (note that $I_+$ and $I_-$ do not commute though);
in fact it is known$^{10}$ that requiring the $p_n$'s to be functions of
$j=h_x$ and its derivatives such that the $I_n$'s commute with $I_+$ and
$I_-$ {\it uniquely} determines the $p_n$'s of the KdV hierarchy. Note
that in our formalism $e^h=q_x$ so (assuming periodic boundary conditions
on $q$) $I_+$ is zero. I strongly suspect (from conformal field theoretic
considerations) that a general set of $p_n$'s
can be obtained by requiring commutation of the $I_n$'s with $I_+$ and
$I(\lambda)=\int dx~e^{-\lambda h}$ for some $\lambda$ (not all $\lambda$'s
will be allowed); but I am unaware of a proof of this statement. In
quantization we will for one purpose
use  ``commutation with $I_+$ and $I_-$''
as the definition of the KdV hamiltonians, and for another purpose use
``commutation with $I_2$'' as the
definition (this is also sufficient to define the other $I_n$'s at the
classical level$^{10}$).

\vskip.2in

\noindent{\bf 3. Quantization}
\smallskip

When we quantize a theory we choose a set of Poisson brackets and elevate
them to the level of operator commutation relations. In quantizing the
theory based on the action $S_0$ we have a choice; either we can treat the
field $q$ as fundamental, in which case we should use the $u$ bracket (4),
as $u$ is the dynamical field, or we can treat the field $h$ as fundamental,
in which case we should use the $j$ bracket (2). But in the latter approach
we should not completely ignore the fact that $S_0$ can be treated as an
action for $q$; this reflects the fact that we can impose a consistent
constraint on the theory defined by $S_0[h]$, namely the constraint
$I_+=\int dx~e^h=0$ (by ``consistent'' in this context I mean that this
constraint is preserved under the dynamics). We will do this.

But first a few words on the standard notions of quantum integrable systems.
A common feature of classical integrable systems is
the existence of at least one Poisson bracket structure and an infinite
number of quantities in involution with respect to this bracket. Given this
situation, we can investigate whether upon elevating the brackets to
operator commutation relations there is still an infinite number of
quantities in involution, with ``classical limit'' (suitably defined)
the hamiltonians of the classical integrable system. If the answer is
positive then we can regard the quantities in involution as conserved
quantities of some operator evolution equation, which we dub the ``quantum''
version of the original classical equation.
Remarkably it seems that there are infinite numbers of conserved quantities
for the quantum KdV equation (quantized using its first$^{10,11,12}$
and second$^{10,11,13}$ brackets), the quantum MKdV equation
(quantized using its second bracket$^{10,13}$),
the quantum NLS equation (quantized using its first$^{14,12}$  and
second$^5$ brackets), and the quantum $SL(N)$ KdV equations
(quantized using its second bracket$^{11,13}$). These are remarkable results
because the ``bihamiltonian'' structure of integrable systems, often
regarded as responsible for the existence of the infinite number of conserved
quantities, is {\it lost} on quantization$^{11}$.

Returning now to the quantization of our action, the first quantization of
$S_0$ proposed above consists of making the $u$ Poisson bracket (4) into an
operator commutation relation. Looking at ref.3 we see that if we write
$$ u=-{{12}\over c}\sum_{n=-\infty}^{\infty}L_ne^{inx} +
     {\textstyle{1\over 2}}  \eqno{(14)}$$
then the modes $L_n$ satisfy a Virasoro algebra with central charge $c$. The
natural choice of Hilbert space is the ``Verma module of the identity'',
i.e. the states
$$ L_{-n_1}L_{-n_2}....L_{-n_r}\vert 0 \rangle,~~~~~~n_1\ge n_2\ge....\ge n_r
        \ge 2 \eqno{(15)}  $$
where $\vert 0 \rangle$ is a vacuum state satisfying
$$ L_n\vert 0 \rangle = 0,~~~~~~n\ge -1 \eqno{(16)}$$
Since this quantum
theory knows nothing of the classical fields $h$ and $j$, we proceed by
defining a quantum analog of $I_2$, namely
$$ {\cal I}_2={1\over 2}\int dx~ (uu)   \eqno{(17)}$$
where the parentheses denote normal ordering. We seek quantum KdV hamiltonians
as operators that commute with ${\cal I}_2$; this is just the quantum KdV
theory of Kupershmidt and Mathieu$^{11}$. It has been proven that an infinite
number of quantum hamiltonians exist$^{13}$.

For the second quantization the fundamental field is $j$; writing
$$ j=\sqrt{6\over c}\sum_{n=-\infty}^{\infty} j_ne^{-inx}   \eqno{(18)}$$
we find the modes $j_n$ satisfy the Heisenberg algebra
$$ [j_n,j_m]=2n\delta_{n,-m}   \eqno{(19)}$$
Without imposing the constraint the natural Hilbert space is the set of states
$$ j_{-n_1}j_{-n_2}....j_{-n_r}\vert 0 \rangle,~~~~~~n_1\ge n_2\ge....\ge n_r
        \ge 1 \eqno{(20)} $$
where here $\vert 0 \rangle$ is a vacuum state satisfying $j_n\vert 0
\rangle=0$, $n\ge 0$. To impose the constraint, the quantum analog of $I_+=0$,
we restrict to states $\vert\psi\rangle$ satisfying
$$ {\cal I}_+\vert\psi\rangle=0  \eqno{(21)}$$
where
$$ {\cal I}_+ = \int dx~:e^h:    \eqno{(22)}$$
In Eq.(22) the colons denote normal ordering and $h=\partial_x^{-1}j$.
Operators in the constrained theory should commute with ${\cal I}_+$ so that
they map physical states to physical states. But before  we work out the
simplest such operator, let us first do some rescalings to make our
formulae appear more like the conformal field theory literature; writing
$$\eqalign{ \phi&=i\sqrt{c\over 6} h \cr
            J   &=\phi_x  }\eqno{(23)}$$
we have
$$\eqalign{ S_0&={1\over{8\pi}}\int dxdt~\phi_x\phi_t \cr
            {\cal I}_+ &= \int dx~:e^{-i\beta\phi}:~,~~~~~~~\beta=
            \sqrt{6\over c} } \eqno{(24)}$$
Following ref.10, we can use conformal field theoretic techniques to
evaluate commutators. We find that the operator $T=-{1\over 4}:J^2:
+i\alpha J_x$ commutes with ${\cal I}_+$ if $\alpha={1\over 2}(\beta-
\beta^{-1})$. $T$ is the analog of $u$ in this quantization of the theory,
and the modes of $T$ satisfy a Virasoro algebra, but with central charge
$\tilde{c}=1-24\alpha^2=13-6(\beta^2+\beta^{-2})=13-c-36c^{-1}$.

For $\beta=\sqrt{m/(m+1)}$, $m=3,4,...$, we obtain in this way
the central charges of the minimal conformal models. In fact what we have
seen here is that quantizing $S_0$, treated
as an action for the constrained field $h$, leads us naturally to
certain features of the Dotsenko-Fateev-Feigin-Fuchs construction for
the minimal models. As explained by Felder$^{15}$, this construction
works because the Hilbert spaces of the minimal models (which are
representation spaces of the Virasoro algebra) can be realised
as the cohomology of a certain operator acting between certain
representation spaces of the Heisenberg algebra (``Fock spaces'').
We have obtained a Lagrangian prescription of a part of this; the states
in our theory are restricted to lie in the kernel of ${\cal I}_+$, which
on the single charge-zero Fock space we have been considering, is Felder's
BRST operator. It might be hoped that a more careful analysis of $S_0$
might lead to a more complete Lagrangian prescription of Felder's work;
in particular in ref.4 I explained why the field $h$ should be regarded as
compactified, and this would motivate us to enlarge the Hilbert space of the
unconstrained theory to include Fock spaces of exactly the charges required.
But it is at the moment not clear to me how the constraint operator
becomes Felder's operator on these  spaces.

Returning to the main subject, we have seen that operators in
the quantum theory of $S_0$ we are now considering must commute with
${\cal I}_+$, and it is therefore natural to seek quantum KdV hamiltonians
in this context by seeking operators that commute both with ${\cal I}_+$
and with ${\cal I}_-\equiv\int dx~:e^{-h}:$~. It is not clear that the
quantum KdV hamiltonians defined this way will coincide with those
defined previously. But it turns out that the second hamiltonian
constructed this way can be written in the form $\int dx~(TT)$ (cf.
Eq.(17)), so the set
of quantum KdV hamiltonians defined here does coincide with those
defined above (up to a replacement of $c$ with $\tilde{c}$). This is a
non-trivial result, that the two definitions of the classical KdV
hamiltonians given at the end of section 2, namely
 ``commutation with $I_+$ and $I_-$'' and ``commutation with $I_2$'',
give rise to the same set of quantum hamiltonians (up to $c\rightarrow
\tilde{c}$)\footnote*{There are several other results we are taking
for granted; for example, on the classical level it is
straightforward to show that the set of objects that commute with $I_2$
generate an abelian algebra, but this is not so clear on the quantum level.
Such issues are discussed in ref.10.};
we will appreciate this result more in the next section.

Finally in this section we note the significance of the quantum KdV
hamiltonians in the minimal models. Defining the quantum KdV hamiltonians
via commutation with ${\cal I}_+$ and ${\cal I}_-$, we see that (for
appropriate values of $\beta$) they are
operators in the minimal model which commute with $\int dx~:e^{-h}:$~; but
$:e^{-h}:$ is exactly the (1,3) primary field, so {\it the
quantum KdV hamiltonians are conserved quantities in the $\Phi_{(1,3)}$
deformations of the minimal models}$^6$.

\vskip.2in

\noindent{\bf 4. An action for the Nonlinear Schr\"odinger Hierarchy}
\smallskip
I will now give the NLS action. In the KdV case, while the action
gave us an interesting perspective, we did not really learn anything new.
In writing the NLS action a) we gain insight into the gauge equivalence
class of the NLS equation (it only takes fragmented knowledge of the class
to write the action, and then it can be used to deduce more), b) on
quantization we see how just as quantum KdV is related to (a deformation
of) the minimal models, similarly quantum NLS is related to (a deformation
of) the parafermion and $SL(2)/U(1)$ coset models, and c) the most
cryptic element in  the bosonization of these models, the form of (one
of the) screening operators is obtained naturally from the classical
theory. I will just give a few details here;
for a more complete discussion see ref.5.

The gauge equivalence class of NLS is specified by giving the Ur-NLS
equations, their second Poisson bracket structure and the group action that
leaves the equations and brackets invariant. Calling the Ur-NLS fields
$S,T$, the equations, brackets and group action are
$$ \eqalign{ T_t&=T_{xx}+2T_xS_x\cr
             S_t&={{2S_xT_{xx}}\over{T_x}}+3S_x^2-S_{xx}  } \eqno{(25)}$$
$$ \pmatrix{ \{S(x),S(y)\} &
             \{S(x),T(y)\} \cr
             \{T(x),S(y)\} &
             \{T(x),T(y)\} \cr}
  = \pmatrix{ 0 &
              -\partial_x^{-1}T_x\partial_x^{-1} \cr
             \partial_x^{-1}T_x\partial_x^{-1} &
             2\partial_x^{-1}T_x\partial_x^{-1}T_x\partial_x^{-1} \cr}
      \delta(x-y) \eqno{(26)}$$
$$ \eqalign{ e^S &\rightarrow \lambda(cT+d)e^S \cr
             T   &\rightarrow {{aT+b}\over{cT+d}} } \eqno{(27)}$$
where in Eq.(27) $ad-bc=1$. The group is $SL(2)\times{\bf R}$.
In the KdV case we had four different variables that were useful, $q,h,j,u$.
$S,T$ are the analogs of $q$, and the analogs of $h,j,u$ are, respectively
the three sets of variables
$$ \eqalign{h&=S\cr
            \bar{h}&=-S+\ln(S_xT_x^{-1})  }
          \eqno{(28)}$$
$$ \eqalign{j&=h_x\cr
            \bar{j}&=\bar{h}_x } \eqno{(29)}$$
$$ \eqalign{A&={\textstyle{1\over 2}}(j-\bar{j}+j_x/j)\cr
            B&=j\bar{j}}\eqno{(30)}$$
The variables $A,B$ are invariant under the full transformation group.
Eq.(25) induces the NLS equation for the quantities $\psi,\bar{\psi}$,
where
$$ \eqalign{\psi&=e^{h-\bar{h}}h_x\cr
            \bar{\psi}&=e^{-(h-\bar{h})}\bar{h}_x} \eqno{(31)}$$
these are invariant under the $SL(2)$ subgroup of the transformation
group; indeed the NLS equations
$$\eqalign{\psi_t&=\psi_{xx}-2\psi^2\bar{\psi}\cr
           \bar{\psi}_t&=-\bar{\psi}_{xx}+2\bar{\psi}^2\psi} \eqno{(32)}$$
display an obvious invariance under $\psi\rightarrow\alpha\psi$
$\bar{\psi}\rightarrow\alpha^{-1}\bar{\psi}$, which is the residual ${\bf R}$.
The NLS action is
$$ S_{NLS}~=~\tilde{k}\int dxdt~h_x\bar{h}_t
          ~+~\sum_{n=1}^{\infty}\lambda_n\int dxdt~p_n[A,B]
                  \eqno{(33)}$$
where the $p_n$'s are certain functionals of $A,B$ and their $x$-derivatives;
one way to define them is to require commutation with
${\cal H}_1\equiv\int dx~e^{h+\bar{h}}$ and
${\cal H}_2\equiv\int dx~h_xe^{-(h+\bar{h})}$,
which play the role of $I_-$ and $I_+$ in the KdV theory. In fact, when
written in terms of $S,T$, and assuming periodic boundary conditions,
${\cal H}_2$ vanishes, just as $I_+$ vanishes in terms of $q$ in KdV theory;
similarly $A,B$ commute with ${\cal H}_2$, just as $u$ commutes with $I_+$
in KdV theory. $S_{NLS}$ gives the $A,B$ hierarchy when varied with
respect to $S,T$, and vice-versa; it gives the $j,\bar{j}$ hierarchy when
varied with respect to $h,\bar{h}$, and vice-versa; it gives the usual
NLS hierarchy when varied with respect to variables $T$ and $S_x/T_x$.

Quantizing $S_{NLS}$ along the lines of the second quantization method
in section 3, we are naturally led to consider states in a Fock space
annihilated by a normal-ordered
version of ${\cal H}_2$, and we have to
seek operators that commute with this constraint operator. It is easy
to find quantized analogs of $\psi,\bar{\psi},B$ (written in terms of
$h,\bar{h}$), and these take exactly the
forms of the fundamental parafermion, its conjugate, and the stress-energy
tensor in the bosonized version of the parafermion and $SL(2)/U(1)$ coset
models. And as I have already said, the normal-ordered version of
${\cal H}_2$ {\it is} the mysterious screening operator in these theories,
which we have now obtained from a simple classical argument.
The quantum NLS hamiltonians, defined by ``commutation with ${\cal H}_1$
and ${\cal H}_2$'' (both normal ordered) can thus be identified as
conserved quantities in a deformation of the parafermion and $SL(2)/U(1)$
coset models by an operator $e^{h+\bar{h}}$, which is just the first ``thermal
operator''. Finally I should mention that
searching for the conserved quantities in the deformed
theories by looking for operators (written in terms of $j,\bar{j}$)
that commute with ${\cal H}_1$
and ${\cal H}_2$ is quite a bit easier than trying to build such operators out
of the quantized $\psi,\bar{\psi}$ fields.

\vskip.2in

\noindent{\bf Acknowledgements}
\smallskip

Useful discussions with Didier Depireux are acknowledged. This work was
supported by the U.S.Department of Energy under grant \#DE-FG02-90ER40542.

\vskip.2in

\noindent{\bf References}
\smallskip

\noindent
\item{1.} From an action one derives both equations of motion {\it and}
Poisson brackets for the fields involved (see ref.4).
It is fairly easy to write an action
which gives the KdV equation and its ``first'' hamiltonian structure, and
this has been done explicitly in (for instance)
L.A.Dickey, {\it Ann.N.Y.Acad.Sci.} {\bf 410} (1983) 301 and
A.Das, {\it Integrable Models}, World Scientific (1989). I am also
informed, by Boris Kupershmidt, that it is known that one can write an
action for the MKdV equation with its ``second'' hamiltonian structure,
but the novel point that I will be stressing here is that this same action
can be viewed as an action for the KdV equation in its ``second''
hamiltonian formulation as well. Note also that from
the work of Dickey just mentioned it is clear such an action exists, but
it is not clear how to write it in a simple form.
\item{2.} For the action mentioned in note 1. for the KdV in its ``first''
hamiltonian formulation this does not seem to be the case.
\item{3.} The first observation of this kind is due to
J.-L. Gervais, {\it Phys.Lett.B} {\bf 160} (1985) 277.
\item{4.} J.Schiff, {\it The KdV Action and Deformed Minimal Models},
     Institute for Advanced Study Preprint IASSNS-HEP-92/28 (revised version).
\item{5.} J.Schiff, {\it The Nonlinear Schr\"odinger Equation and Conserved
      Quantities in the Deformed Parafermion and SL(2,R)/U(1) Coset Models},
      Institute for Advanced Study Preprint IASSNS-HEP-92/57.
\item{6.} A.B.Zamolodchikov, {\it Adv.Stud.in Pure Math.} {\bf 19}
      (1989) 641.
\item{7.} It is often stated that the MKdV equation has only one hamiltonian
    structure. In fact, as shown in F.Magri, {\it J.Math.Phys.} {\bf 19}
    (1978) 1156, it has two hamiltonian structures, one local and one
   non-local. I call the local one ``second'' and the non-local one ``first''.
\item{8.} G.Wilson, {\it Phys.Lett.A} {\bf 132} (1988) 45;
       {\it Quart.J.Math.Oxford} {\bf 42} (1991) 227;
      {\it Nonlinearity} {\bf 5} (1992) 109; and in {\it Hamiltonian
     Systems, Transformation Groups and Spectral Transform Methods},
     ed. J.Harnad and J.E.Marsden, CRM (1990).
\item{9.} A.M.Polyakov, {\it Mod.Phys.Lett.A} {\bf 2} (1987) 893;
        M.Bershadsky and H.Ooguri, {\it Comm.Math.Phys.} {\bf 126}
       (1989) 49 and references therein.
\item{10.} R.Sasaki and I.Yamanaka, {\it Comm.Math.Phys.} {\bf 108}
    (1987) 691; {\it Adv.Stud.in Pure Math.} {\bf 16} (1988) 271.
\item{11.} B.A.Kupershmidt and P.Mathieu, {\it Phys.Lett.B} {\bf 227}
              (1989) 245.
\item{12.} M.D.Freeman and P.West, {\it On the quantum KP hierarchy and its
    relation to the non-linear Schr\"odinger equation}, King's College
    preprint KCL-TH-92-2, to appear in {\it Phys.Lett.B}.
\item{13.} B.Feigin and E.Frenkel, {\it Phys.Lett.B} {\bf 276} (1992) 79.
\item{14.} M.Omote, M.Sakagami, R.Sasaki and  I.Yamanaka, {\it Phys.Rev.D}
      {\bf 35} (1987) 2423.
\item{15.} G.Felder, {\it Nucl.Phys.B} {\bf 317} (1989) 215.

\bye